\documentclass[preprint,tightenlines,a4paper,amsmath,amssymb,superscriptaddress,nofootinbib,
showpacs]{revtex4-1}

\usepackage{bm}

\begin{document}

\title{Hyperfine Effects in Ionic Orbital Electron Capture}

\author{M.A. Go\~ni}

\affiliation
{
Theoretical Physics Department,
 Euskal Herriko Unibersitatea,
 644 Posta Kutxatila, 48080 Bilbo, Spain
}

\date{\today}

\begin{abstract} 
The K-orbital electron capture in ions with one or two electrons is analized for a general allowed nuclear transition. For ionic hyperfine states the angular neutrino distribution and the electron capture rate are given in terms of  nuclear matrix elements. 
A possible application towards the determination of neutrino parameters is outlined.
\end{abstract}



\maketitle

\section*{Introduction}
Significant modifications of weak nuclear rates due to the influence of atomic electrons have been predicted and their importance emphasized. J.N.Bahcall~\cite{Bah} developed a general 
discussion and treatment of the correlations and rates of bound  beta decay processes for arbitrary electronic configurations and their effects on
the behavior of hot stellar plasmas. 
Recently, an experiment at GSI has reported~\cite{Otsh} the measurement of the ratio of bound state to continuum state total beta decay  rates   for the case of bare $1/2^{+}$ ${}^{207}$Tl decaying into the ground state of $1/2^{-}$ ${}^{207}$Pb. It is a nuclear first forbidden transition  and the ratio is  in excellent agreement with the theory~\cite{Taka},~\cite{Fab}.\\

New facilities allow electronic capture (EC) and $\beta^{+}$ studies on highly ionized atoms,  combining novel experimental tools which use high energy accelerators, in-flight separators and heavy ions storage rings.
An experiment performed recently at GSI~\cite{1Lit} has studied the radioactive ionic decays of $1^{+}$ ${}^{140}$Pr (Z=59)  with 0,1,2 K-orbital electrons into $0^{+}$ ${}^{140}$Ce (Z=58). The Lorentz factor  $\gamma$ is 1.43 and the results reported in~\cite{1Lit}  are EC and $\beta^{+}$ rates in the rest frame of ions. In this  new kind of experiments the atomic contribution is well known, permitting cleaner
measurements of the  weak  nuclear parameters. EC experiments on neutral atoms were reviewed in ref.~\cite{Bam}.\\

The hyperfine structure due to the coupling of the nuclear spin $I$ and the angular momentum $1/2$ of the K-electron,  giving a total angular momentum $F=I\pm 1/2$  ($F_{\pm}$), is fundamental in order to understand the EC results.
 The ground state (gs) of  ${}^{140}$Pr${}^{58+}$ has $F=1/2$, as follows from its positive magnetic moment $\mu$~\cite{Sha}, whereas the weak decay of the excited $F=3/2$ state is forbidden in the allowed aproximation. As the relaxation time for the upper hyperfine state is much shorter than the cooling time,  the ions are dominantly stored with $F=1/2$  in the experiment considered in~\cite{1Lit}.\\

The importance and origin of the influence of the hyperfine structure on the decay rates  was   clearly showed  by Folan and Tsifrinovich~\cite{Fol} in their seminal analysis of EC decays of H-like atoms. Among other results, they showed that, in a spin $1/2$ mirror transition ${}^{31}$S$\rightarrow $ ${}^{31}$P, the rate for the  $F=0$ states is 340 times larger than the rate for  states with $F=1$. They also pointed out the possibility of observing  this kind of  effects with trapping techniques of cold ions.\\

More recently, a theoretical analysis  of  general EC hyperfine rates has been given in~\cite{Pat}, where the ratio of the decay rates  for  helium -like   and hidrogen -like ions is calculated explicitly as a  function of the  nucleus spin $I$ in the allowed aproximation, and suitable candidates for an experimental confirmation are discussed. 
 A more refined analysis of K-shell  EC and $\beta^{+}$ weak decays rates  of $^{140}$ Pr $^{58+}$and$ ^{140}$ Pr$^{ 57+}$, which takes into account the nuclear size and   uses  relativistic electron wave functions, has been presented in~\cite{1Yva,2Yva}. The ratios of the decay rates computed with  these refinements    agree with experimental results  with an accuracy better than 3 percent. \\
 
  Experiments at GSI have also measured time dependence of the electron capture rate of H-like ${}^{140}$Pr${}^{58+}$, ${}^{142}$Pm${}^{60+}$ and ${}^{122}$I${}^{52+}$ ions~\cite{2Lit,Win,Kie}. They  have found an A-dependent modulation in the capture rate  $dN_{EC}(t)/dt$ that has been interpreted as due to neutrino oscilations~\cite{3Yva}, although  this interpretation is still controversial, see for instance~\cite{4Yva,Giu}.\\
  
 In a different line of developments,  EC  experiments have been proposed,~\cite{Sat1,Ber1}, as sources of monoenergetic neutrino beams, aiming at the determination of the neutrino mixing angle $\theta _{13}$, and the CP violation parameter $\delta _{CP}$. The experiments would use extremely high energy heavy ions ($\gamma $ $\approx$ 10$^{2}$-10$^{3}$)~\cite{Zuc} that capture one electron and produce a  line neutrino spectrum which is  rotationally symmetric in the CM frame.  Neutrino mixing would change the nature of the neutrinos detected  at a distance $L$. 
The fluxes of $\nu_{\mu}$ are predicted by well known methods, see for instance~\cite{Ber2} and references therein. Currently, a systematic study of EC in rare-Earth nuclei, relevant for neutrino beams, is being carried out~\cite{Alg}.\\

 In this note a general analysis of allowed K-electronic capture, still lacking in the literature as far as we know, is presented  and an application to the measurement of  neutrino mixing parameters using polarized ions is outlined.
 The allowed transitions from the mother nucleus $i$ into the daughter $f$ give the selection rules $I \rightarrow I'=I-1$ ($F_{-}$),  $I'=I$ ($F_\pm$) and $I'=I+1$ ($F_+$).The cases $I'= I \pm 1$ are pure Gamow-Teller, whereas for $I'= I$  both Fermi and Gamow-Teller contribute. In all cases, the initial and final parities are the same ($\pi _{i}= \pi _{f }$). In what follows,  we consider separately the EC decays of H-like and He-like ions. 

\section*{H-like EC}
Let us consider the transition amplitude for a K-EC process from an initial hyperfine $|FM >$ state to a final state where the nucleus spin  and $z$-component are $I',m'$ respectively  and the left neutrino has momentum $\vec q _{\nu}$= $E_{\nu} \vec n_{\nu}$. This amplitude will be denoted as $T_{FM\rightarrow m'\nu}$.\\
The general structure of gs H-like transition amplitudes $T_{FM\rightarrow m'\nu}$ in terms of   reduced   angular momentum Wigner-Eckart amplitudes  $\overline{T}$ is
\begin{eqnarray} 
T_{FM\rightarrow m'\nu} &=& a_{+} (\vec n _{\nu})\delta _{m',M'-1/2}
\left[\sqrt{\frac{I'+M'+1/2}{2I'+1}}<F'_{+}M'|T|FM> \right. \nonumber \\ 
&& -\left.\sqrt{\frac{I'-M'+1/2}{2I'+1}}<F'_{-}M'|T|FM>\right] \nonumber \\
&&+ a_{-}(\vec n _{\nu})\delta _{m',M'+1/2} 
\left[\sqrt{\frac{I'-M'+1/2}{2I'+1}}<F'_{+}M'|T|FM> \right. \nonumber \\
&&+ \left.\sqrt{\frac{I'+M'+1/2}{2I'+1}}<F'_{-}M'|T|FM>\right] .      
\end{eqnarray}
where the fundamental property
\begin{equation} 
<F'M'|T|FM>=\delta _{F',F}\delta _{M,M'} \overline{T}_{F\rightarrow F'}
\end{equation}
incorporates  angular momentum conservation and rotational invariance at the outset and  
the left handed  neutrino spin wave function is given by
\begin{eqnarray}
|\nu_{L}>&=& a_{+}(\vec n _{\nu})^{*}|\uparrow> + a_{-}(\vec n _{\nu})^{*} |\downarrow>  \nonumber \\
         &=& -\sin (\theta /2) e^{-i\phi} |\uparrow> + \cos (\theta /2) |\downarrow> . 
\end{eqnarray}

The reduced $\overline{T}$ are obtained by calculating suitable   weak interaction $S$-matrix elements at first order, using the standard $\Delta S=0$ piece of the Hamiltonian  with coupling constant $G_{F}V_{ud}/\sqrt{2}$ and  renormalized  neutron beta decay axial coupling $g_{A}$~\cite{Ams}. The matrix elements factorize into a lepton factor that can be  computed explicitly and the nuclear matrix elements of the Vector $V^{\mu}$ and Axial $A^{\mu}$ weak hadronic currents.  In the allowed aproximation these nuclear matrix elements depend only on two caracteristic constants $\mathcal{M}_{\sigma,F }(see $~\cite{des} and the Appendix), that one should obtain by using a good nuclear model.\\

The results for the reduced $\overline{T}$, with the weak coupling constants and the value of the electron wave function omitted, are collected in the Appendix.
The cases $I'=I\pm 1$ are pure Gamow-Teller, whereas in the case $I'=I$ both Gamow-Teller and Fermi contribute so that  the hiperfine rates depend on $\mathcal{M}_{F}^{2}$, $\mathcal{M}_{\sigma }^{2}$ and the interference term $\mathcal{M}_{F}\mathcal{M}_{\sigma}$. As expected, the interference disappears when the initial spin  ion is not  polarized ~\cite{Bam,des}. The $I = 1/2$ results in~\cite{Fol}  are easily recovered.\\

 For fixed $M$, $|T _{FM\rightarrow m'\nu}|^{2}$  depends on both   the final neutrino direction and the polarization of the final nucleus. Upon summing over the unobserved polarization of the daughter nucleus
 the angular distribution exhibits a characteristic $\cos\theta$ linear distribution.
For instance in the case  $ I'$=$I-$1 (the remaining  cases  are collected in the Appendix) we find 
\begin{equation}
\sum _{m' }|T _{F_{- }M\rightarrow m'\nu}|^{2} =\left(
    \frac{1}{2}- \frac{M}{2I-1}\cos\theta \right)|\overline{T}_{F_-}|^{2} .
\end{equation}

The capture rate is $M$-independent when one sums over both final nuclear polarization and neutrino momentum (rotational invariance)

\begin{equation} 
\int d\Omega _{\nu}\sum_{m'} |T _{FM\rightarrow m'\nu}  |^{2}= 2 \pi |\overline{T}|^{2} 
\end{equation}  
and the hyperfine rate $W$ is 
\begin{equation} 
\label{eq:W}
W=\frac{(G_{F}V_{ud})^{2}}{\pi}(g_{A}\mathcal{M}_{\sigma})^{2}|\varphi_{0}|^{2}Q_{\nu}^{2}\frac{2I+1}{2I}
\end{equation}
where $Q_{\nu}$ is the neutrino energy. The two body phase space originates the $Q_{\nu}^{2}$ factor.\\

 In the K-EC $1^{+}\rightarrow 0^{+}$ decay for $^{140}$Pr$^{58+}$, or in a $0^{+}\rightarrow 1^{+}$ transition~\cite{Ber1}, the use of an  effective  interaction hamiltonian density in terms of effective  relativistic nuclear spin 1, 0  fields $H_{\mu},\phi$ 
\begin{equation}
g\overline{\nu}(1+\gamma _{5})\gamma ^{\mu}eH_{\mu}\phi +h.c.  
\end{equation}
provides a quick way to get our results. The amplitude from initial $I=1,I_{z}=m, S_{z}=s$ to $m'=0, \nu$ is
\begin{equation} 
T _{ms\rightarrow 0\nu}\propto \chi_{ L}^{+}\vec \sigma \chi _{s}\vec \epsilon _{m}
\end{equation}
where $\vec \epsilon _{m}, m=\pm1,0$, are the  spin-$1$ states of the mother nucleus and $\chi_{s,L}$ are 
the Pauli  spin  function of the captured electron  and the final neutrino respectively. 
Elaborating the hyperfine amplitudes turn out to be \\
\begin{equation} 
T_{F=1/2,\lambda} \propto -\sqrt{3} \chi _{L}^{+}\chi _{\lambda},\qquad T _{F=3/2,\lambda}=0  
\end{equation}
in agreement~\cite{F1} with the general results in the  Appendix.

\section*{He-like EC}
For ions with two K electrons in  the ground state, S=0, the capture process  
changes the initial $|(ee)_{S=0}I m >$ state  into the state $|I 'm'\nu e_{s'}>$, with a daughter nucleus, one neutrino  and  one bound spectator electron. The transition amplitude is
\begin{equation}   
T^{He}_{(ee)_{gs}Im \rightarrow I 'm'\nu s'}= (T^{H} _{(e\uparrow)_{gs}Im \rightarrow I 'm'\nu}) \delta _{s'\downarrow}-(T^{H} _{(e\downarrow)_{gs}Im \rightarrow I 'm'\nu}) \delta _{s'\uparrow} 
\end{equation}
and the transition probability after summing over final electron 
spin W$_{(ee)_{gs} m  \rightarrow  m'\nu}$ is thus given by
\begin{equation}
W_{(ee)_{gs} m  \rightarrow  m'\nu}=\left|(T^{H} _{(e\uparrow)Im \rightarrow I 'm'\nu})\right|^{2}+\left|(T^{H} _{(e\downarrow)Im \rightarrow I 'm'\nu})\right|^{2}
\end{equation}
In terms of the hyperfine amplitudes  
\begin{eqnarray}
(2I+1)&&W_{(ee)_{gs}m \rightarrow  m'\nu}=  \nonumber \\ 
&&
(I+m+1)\left|T^{H} _{F_{+},M=m+1/2\rightarrow I 'm'\nu}\right|^{2}+
(I-m)\left|T^{H} _{F-,M=m+1/2\rightarrow I 'm'\nu}\right|^{2} \nonumber \\
&&-2\sqrt{(I+1/2)^{2}-M^{2}} \mathrm{Re}[T^{H}_{F+,M=m+1/2\rightarrow I 'm'\nu}(T^{H}_{F-,M=m+1/2\rightarrow I 'm'\nu})^{*}] \nonumber \\
&&+(I-m+1)\left|T^{H} _{F_{+},M=m-1/2\rightarrow I 'm'\nu}\right|^{2}
+(I+m)\left|T^{H} _{F-,M=m-1/2\rightarrow I 'm'\nu}\right|^{2} \nonumber \\
&&+ 2\sqrt{(I+1/2)^{2}-M^{2}}\mathrm{Re}(T^{H}_{F+,M=m-1/2\rightarrow I 'm'\nu}(T^{H}_{F-,M=m-1/2\rightarrow I 'm'\nu})^{*})  
\end{eqnarray}

Upon summing and integrating over final $m', \nu$ the interference term disappears and the RHS of (11) becomes
\begin{eqnarray}
&&(I+m+1)\sum_{m' \nu}\left|T^{H}_{F_{+},M=m+1/2\rightarrow I 'm'\nu}\right|^{2}+
(I-m+1)\sum_{m' \nu}\left|T^{H} _{F_{+},M=m-1/2\rightarrow I 'm'\nu}\right|^{2} \nonumber\\
&& +(I-m)\sum _{m' \nu}\left|T^{H}_{F_{-},M=m+1/2\rightarrow I 'm'\nu}\right|^{2} .
\end{eqnarray} 
\\
Therefore the total H, He-like  $W$ rates are related by
 \begin{equation}
 W^{He}= 2\frac{I+1}{2I+1}W ^{H}_{(+)}+2\frac{I}{2I+1}W ^{H}_{(-)}
 \end{equation}
 This relation was first obtained in~\cite{Pat}. For $^{140}$Pr it gives good agreement with experiment~\cite{1Lit} once the important corrections due to  relativistic Coulomb effects are taken into account~\cite{1Yva,2Yva}.\\ 
 
In the case $1^{+}\rightarrow 0^{+}$ one obtains  
 \begin{equation}
 3W_{(ee)_{gs} m  \rightarrow  0\nu}=
 (1-m)\left|T^{H}_{1/2,M=m+1/2\rightarrow  0\nu}\right|^{2}+
 (1+m)\left|T^{H}_{1/2,M=m-1/2\rightarrow  0\nu}\right|^{2}
 \end{equation}
and  therefore 
 \begin{equation}
 W_{(ee)_{gs} m  \rightarrow  0\nu}\propto (1-m\cos\theta) ,\quad m=1, 0, -1 .
 \end{equation}
 
  As required by rotational invariance, the magnetic $m$-number dependence vanishes after integrating over the neutrino directions. This dependence disagrees~\cite{F2}  with the results of~\cite{1Yva,2Yva}.
 In the general case $I\rightarrow I-1$ the  neutrino angular  dependence is given by
 \begin{equation}
  \frac{I-m \cos\theta }{I},\qquad m =I, I-1,..., -I .
  \end{equation}

\section*{EC and Neutrino Parameters}

EC could be useful in order to fix the values of the yet unknown neutrino mixing parameters 
$\theta_{13}, \delta _{CP}$~\cite{Ber1,Ber2}.\\

Monoenergetic -- in the EC rest frame --  pure electronic neutrinos  are detected as $\nu _{\mu}$ in a long baseline  neutrino experiment. Accelerated stored high gamma ions have been proposed as suitable neutrino sources (Zucchelli~\cite{Zuc}).\\

J. Sato~\cite{Sat1} and  M. Rolinec and J. Sato~\cite{Sat2} have investigated the physical potential for neutrinos coming from the EC process e$^{-}$ $^{110}$Sn$\rightarrow$ $^{110}$ In $^{*}$ $\nu _{e}$, $Q = 267$ kev,  which is an  allowed $0^{+}\rightarrow 1^{+}$ nuclear transition. 
H-like ions would be assembled by using two equal $\gamma$ paralell beams of bare nuclei and electrons that would be  captured in fly. As $Q$ is small the lifetime should be large   
(recall Eq.(\ref{eq:W}), experimentally $\tau$ = 4.11h) and therefore the emited neutrino flux would be low. The proposed  Setups have baseline $L = 250$, $600$ km and $\gamma = 900-2000$.\\

J. Bernabeu and collaborators have independently proposed~\cite{Ber1} EC neutrino factories and thoroughly studied the fenomenology and viability of  EC  neutrino experiments aiming to measure the CP violating phase $\delta _{CP}$. Recently~\cite{Ber2} they have   proposed an hybrid beta decay and EC Setup  using $^{156}$ Yb ions that decay 
38\% via EC with a $\nu$-energy of 3.46 Mev and 52\% via $\beta$ $^{+}$ with  end neutrino energy of $2.44$ Mev. The daughter nucleus  $^{156}$Tm$^{*}$ is an excited $1^{+}$ giant Gamow-Teller resonance state so that  the halflife, $t_{1/2}=26.1$ seconds, is short enough to allow EC in the decay ring.\\

One should note that the use of polarized H-like ions would produce a neutrino flux depence inside the detector that could be useful in order to disentangle the values of neutrino mixing parameters. In the case of a 0$^{+}\rightarrow$ 1$^{+}$ nuclear transition with capture of one  K-electron
the neutrino angular distribution  from  ions with polarization vector $\vec P$ = $< \vec \sigma >$ is

\begin{equation}
W(\vec P,\vec n _{\nu} ) =\frac{1}{2}+\frac{1}{3}\vec P\cdot\vec n _{\nu},\qquad |\vec P |= p 
\end{equation}
With  $\vec P$ in the ion beam direction, the   $\nu$-distributions in the ion rest frame (rf) show  a caracteristic parity violating  linear $\cos \theta _{rf}$ dependence

\begin{equation}
W(\theta _{rf}, p ) =\frac{1}{2}(  1 \pm \frac{p }{3}\cos \theta _{rf}) 
\end{equation}

 This angular modulation is observable in the Rolinec, Sato proposal, Setup II, with a large Water Cherenkov detector of radius R = 100 m, base L = 250 km. With this geometry $\tan  \theta _{max}$ = $\frac{R}{L}$ and, for a neutrino emited at right angle $\theta _{rf} = \frac{\pi}{2}$ to be detected,  a  boost
\begin {equation} 
 \underline{\gamma}= \frac{1}{\sin \theta _{max}} = 2500, \qquad 
 \end{equation} 
is  required. Rolinec and  Sato take $\gamma$ = 2000 and therefore  $\cos (\theta _{rf })_{max}$ = .22.
The flux of $\nu_{\mu}$ in points  inside the  detector would depend now, in a known way, also on  the ion polarization. The requirements to achieve such a performance, i.e., very high $\gamma$, large number of isotope decays  per year, very low beam divergences for the stored isotopes~\cite{Sat2}, together with  the use of  very high $\gamma$ polarized electron beams  are extremely demanding.


\newpage

\section*{ACKNOWLEDGMENTS}

I wish to thank  Jos\'e Bernabeu for information on  neutrino factories and to Elvira Moya de Guerra for sharing   her knowledge on weak nuclear properties. I am indebted  to Amalio F. Pacheco for clarifications, encouragement and very useful comments on the manuscript, and to Juan L. Ma\~nes and Manuel Valle for invaluable criticism and suggestions.\\
\begin{center}
This work is dedicated to Prof. Julio Abad, in Memoriam.
\end{center}

\section*{APPENDIX }

\subsection{DEFINITION OF WEAK NUCLEAR  PARAMETERS $\mathcal{M}, A$}
Let $\Psi_{i,f}$ be the  wave functions of the initial, final nuclei at rest. The operators asociated to the allowed transitions are $1 \tau_{-}, \sigma_{m}\tau_{-}$ summed over   nucleons, 
where $\tau_{-}$ is the Isospin lowering matrix converting a proton into a neutron and $\sigma_{m}$ are the standard spherical components of the Pauli $\vec \sigma$ matrices. The nuclear matrix elements $\mathcal{M}_{F,\sigma }$
are defined as follows
\begin{eqnarray*}
 <\Psi(I'm')_{f}\left| 1 \right|\Psi(Im)_{i}> &\equiv& \mathcal{M}_{F}\delta _{m},_{m'}\delta _{I,I'} , \\
 <\Psi(I'm')_{f}\left| \sigma _{n}\right|\Psi(Im)_{i}> &\equiv&  \mathcal{M}_{\sigma }
 \sqrt{\frac{2I+1}{2I'+1}}  <I1mn|I'm'>,\qquad (n=\pm1,0). 
\end{eqnarray*}

\subsection{HYPERFINE  AMPLITUDES}

In that follows we use the definitions
\begin{eqnarray*}
A_{F}&=& \mathcal{M}_{F}, \\ 
A_{\sigma} &=&-g_{A} \mathcal{M}_{\sigma }\sqrt{\frac{2I+1}{2I'+1}} .
\end{eqnarray*}

\subsubsection{$I'=I-1$ CASE}

\[T _{F_{- }M\rightarrow m'\nu} =\left[a_{+}(\vec n _{\nu})\delta _{m',M-1/2}\sqrt{\frac{I+M-1/2}{2I-1}}
+a_{-}(\vec n _{\nu})\delta _{m',M+1/2}\sqrt{\frac{I-M-1/2}{2I-1}}\right] \overline{T}_{F_{-}}\]
\\
\[\overline{T}_{F_-}=\overline{T}_{F_{-}\rightarrow F'_{+}}=(A_{\sigma} )\sqrt{\frac{2I-1}{I}}\]
\\
\subsubsection{ $I'=I$ CASE}

\[T _{F_\pm M\rightarrow m'\nu} =\left[a_{+}(\vec n _{\nu})\delta _{m',M-1/2}(\pm \sqrt{\frac{I\pm M+1/2}{2I+1}})
+a_{-}(\vec n _{\nu})\delta _{m',M+1/2} \sqrt{\frac{I\mp M+1/2}{2I+1}}\right] \overline{T}_{F_\pm}\]\\

\[ \overline{T}_{F_+}=\overline{T}_{F_{+}\rightarrow F'_{+}}=A_{F}+A_{\sigma}\sqrt{\frac{I}{I+1}},\qquad \overline{T}_{F_-}=\overline{T}_{F_{-}\rightarrow F'_{-}}=A_{F}-A_{\sigma}\sqrt{\frac{I+1}{I}} \]

\subsubsection{ $I'=I+1$ CASE}

\[T _{F_+ M\rightarrow m'\nu} =\left[-a_{+}(\vec n _{\nu})\delta _{m',M-1/2}\sqrt{\frac{I-M+3/2}{2I+3}}
+a_{-}(\vec n _{\nu})\delta _{m',M+1/2}\sqrt{\frac{I+M+3/2}{2I+3}}\right] \overline{T}_{F_{+}}\]
\\
\[\overline{T}_{F_+}=\overline{T}_{F_{+}\rightarrow F'_{-}}=-(A_{\sigma})\sqrt{\frac{2I+3}{I+1}} \]
\subsection{NEUTRINO ANGULAR DISTRIBUTIONS, FIXED $M$}

\subsubsection{$I'=I-1$}

\[\sum _{m' }|T _{F_{- }M\rightarrow m'\nu}|^{2} =\left(\frac{1}{2}- \frac{M}{2I-1}\cos\theta 
\right)|\overline{T}_{F-}|^{2} \]

\subsubsection{$I'=I$}

\[\sum_{m' }|T _{F_{\pm} M\rightarrow m'\nu}|^{2} = \left(\frac{1}{2}\mp \frac{M}{2I+1}\cos\theta \right)|\overline{T}_{F_{\pm}}|^{2}  \]

\subsubsection{$I'=I+1$}

\[\sum_{m' }|T _{F_+ M\rightarrow m'\nu}|^{2} = \left(\frac{1}{2}+ \frac{M}{2I+3}\cos\theta \right)|\overline{T}_{F+}|^{2}\]


\begin{thebibliography}{99}
\bibitem {F1}
Note that the  hyperfine amplitudes $T_{F=1/2,M}$ are not vanishing for $M = \pm \frac{1}{2}$. This disagrees with the amplitudes reported in~\cite{1Yva,2Yva}. The discrepancy may be due to the fact that the amplitudes in~\cite{1Yva,2Yva} seem to be referred to a frame where the  variable final neutrino direction is taken as quantization axis. This does not affect  the rates, which are computed by  averaging over initial  and summing over final spin components.
\end{thebibliography}
\end{document}